\documentclass[a4paper,12pt]{article}

\usepackage{graphicx}
\usepackage{array}
\usepackage{amsmath}
\usepackage{amssymb}
\usepackage{latexsym}
\usepackage{subfigure}
\usepackage{color}

\date{empty}
\begin{document}
\begin{titlepage}
\null
\begin{flushright}
TIT/HEP-609 \\
Mar., 2011
\end{flushright}
\vskip 1.0cm
\begin{center}
  {\Large \bf R-symmetry Breaking and O'Raifeartaigh Model \\
\vskip 0.3cm
with Global Symmetries at Finite Temperature}
\vskip 2.0cm
\normalsize
\renewcommand\thefootnote{\alph{footnote}}

{\large
Masato Arai$^{\dagger}$\footnote{masato.arai(at)utef.cvut.cz},
Yoshishige Kobayashi$^\ddagger $\footnote{yosh(at)th.phys.titech.ac.jp}
and Shin Sasaki$^\ddagger$\footnote{shin-s(at)th.phys.titech.ac.jp}
}
\vskip 0.5cm
  {\it
  $^\ddagger$Institute of Experimental and Applied Physics \\
  Czech Technical University in Prague \\
  Horsk\'a 3a/22, 128 00 Prague 2, Czech Republic \\
  \vskip 0.5cm
  $^\ddagger$Department of Physics, Tokyo Institute of Technology \\
  Tokyo 152-8551, Japan
}
\vskip 1.5cm
\begin{abstract}
We investigate finite temperature effects in O'Raifeartaigh models with 
global symmetries which exhibit supersymmetry breaking at a meta-stable
 vacuum accompanied by $U(1)_R$ breaking.
The pseudo moduli field is stabilized at one-loop order 
at zero temperature within some coupling parameter region.
We analyze 
the behavior of the parameter space according to non-zero temperatures and
find that the parameter region which allows $U(1)_R$ breaking
is considerably extended at sufficiently low temperature,
even though it shrinks down at high temperature as expected.
 We also 
discuss the thermal history of the meta-stable supersymmetry breaking vacuum. 
\vskip 0.5cm
\end{abstract}
\end{center}

\end{titlepage}

\newpage
\setcounter{footnote}{0}
\renewcommand\thefootnote{\arabic{footnote}}
\pagenumbering{arabic}
\section{Introduction}
No one doubts that supersymmetry breaking has been the long-standing problem
in high energy particle physics and the problem is 
closely related to the existence of global R-symmetry. 
The relation between supersymmetry breaking and global R-symmetry has
been studied after its connection was established in 
\cite{NeSe}. 
The authors of \cite{NeSe} showed that in order to break supersymmetry
at a global minimum of the potential in generic models, there must
exist global $U(1)_R$ symmetry. However the $U(1)_R$ symmetry should be
broken at the supersymmetry breaking vacua otherwise gauginos can not
have non-zero masses that are phenomenologically required. 

A possible solution to this puzzle is to consider a local but not global
minimum of the potential to break supersymmetry. 
In \cite{InSeSh}, Intriligator, Seiberg and Shih studied 
low-energy effective potential of an $\mathcal{N} = 1$ supersymmetric
QCD (ISS model) which exhibits supersymmetry breaking at a local vacuum. 
This vacuum is meta-stable but can be long-lived, compared to the life
time of the universe. 

Other models with supersymmetry breaking at local vacua have been
studied. It has been shown that O'Raifeartaigh models \cite{Or} 
with fields that have R-charges different from 0 or 2 can have a meta-stable
supersymmetry breaking vacuum without $U(1)_R$ symmetry \cite{Sh}. 
In these models, supersymmetry is broken at tree-level while the
$U(1)_R$ symmetry is broken at one-loop order. Models with tree-level
R-symmetry breaking were investigated in \cite{KoSh}.

It is useful to generalize these models to ones with global
symmetries \cite{Fe}.
Such an extension is important since if global symmetries are gauged 
as gauge groups of the Standard Model, supersymmetry breaking is mediated to 
the visible sector through direct gauge mediation mechanisms 
\footnote{
However, ISS model and models as in \cite{Fe} lead to suppressed gaugino 
mass which is phenomenologically unacceptable \cite{KoSh}. Deformed ISS 
models have been considered in, for instance, \cite{DiMa}.}.
These observations open up more possibilities for constructing 
phenomenologically viable models (for example see \cite{MuNo}) and 
the realization of long-lived local vacua in field theories provides a 
good example to study the string landscape problem \cite{KaKaLiTr}.

It is also interesting to study the thermal effects of these local vacua. 
The thermal effects generically modify the structure of the vacua and change
the view of the landscape. 
Since we believe that supersymmetry is broken at early universe after the 
inflation, it is useful to analyze the thermal history of the hidden sector. 
The thermal history of supersymmetric models with supersymmetry breaking
has been investigated in the context of ISS model 
\cite{AbChJaKh1, CrFoWa, FiKaKrMaTo, AbJaKh} and the O'Raifeartaigh type model
without $U(1)_R$ symmetry \cite{MoSc}.
The thermal effects of a broad class of models of gauge mediation were also considered \cite{Ka}.

In this paper we study effects of finite temperature on 
the O'Raifeartaigh type model with global symmetry considered 
in \cite{Fe}. Classically this model has a flat direction of local extrema 
and a supersymmetric runaway vacuum. 
One loop corrections stabilize the flat direction, leading to a spontaneous
$U(1)_R$ breaking vacuum in some regions of parameters in the theory. 
We take account of finite temperature effects and show that such 
parameter regions shrink down as temperature increases.
We also show that parameter regions which are not allowed at zero temperature
cause $U(1)_R$ breaking at low temperature.
We also address the finite temperature effects along the runaway direction and
discuss the thermal evolution of a local vacuum.

The organization of this paper is as follows. 
In the next section, we introduce the model and discuss the 
supersymmetry breaking vacuum and the runaway direction at zero temperature. 
In section 3, we study finite temperature effects of the model. Especially we 
investigate behavior of the parameter region according to temperatures. 
In section 4, we discuss the thermal history of the local vacuum
focusing on a specific path to the runaway direction. 
Section 5 is conclusion and discussions.
In appendix A, analytic expressions of the effective potential in high
and low temperature expansions are shown.
In appendix B, we give explicit expression of general mass matrices. 

\section{The model and supersymmetry breaking}
In this section, we provide a brief overview of the generalized 
O'Raifeartaigh model that exhibits supersymmetry breaking without
$U(1)_R$ symmetry \cite{Sh}. We then introduce the model with global
symmetry and show the result obtained in \cite{Fe} at zero temperature.

Consider O'Raifeartaigh models with the following superpotential
\begin{eqnarray}
W &=& f X + \frac{1}{2} (M^{ij} + X N^{ij}) \Phi^i \Phi^j, 
\label{OR} \\
  & & \det M \not= 0, \ \det(M + X N) \not=0, \label{det}
\end{eqnarray}
where $M,N$ are complex symmetric matrices, $f$ is a real parameter 
and $X, \Phi^i$ are $\mathcal{N} = 1$ chiral superfields. 
We assume that the K\"ahler potential for $\Phi^i, X$ is canonical. 
The condition of $U(1)_R$ charge for each field is $R(X)=2$ and the
other fields satisfy
\begin{eqnarray}
\begin{aligned}
R(\Phi^i)+R(\Phi^j)=2 & \quad \mbox{for}~M^{ij}\neq 0, \\
R(\Phi^i)+R(\Phi^j)=0 & \quad \mbox{for}~N^{ij}\neq 0.
\end{aligned}
\end{eqnarray}
The supersymmetric vacuum conditions are given by 
\begin{eqnarray}
 0=f+{1 \over 2}N^{ij}\phi^i\phi^j,\quad 0=(M^{ij}+N^{ij}X)\phi^j. \label{cond}
\end{eqnarray}
Here $\phi^i$ and $X$ are the lowest component of the superfields 
$\Phi^i$ and $X$.
Due to the condition (\ref{det}), 
the equations in \eqref{cond}
 can not be solved simultaneously and
the model breaks supersymmetry at an
extremum of the potential:
\begin{eqnarray}
\phi^i = 0, \quad X:\mathrm{arbitrary}. 
\label{local_vacuum}
\end{eqnarray}
This solution leaves a flat direction for the $X$ field,
which is called pseudo modulus. 
In order that this extremum is a local vacuum, there must not exist 
tachyonic directions.
The bosonic and fermionic mass matrices at this extremum are given by
\begin{eqnarray}
m_B^2 &=& 
\left(
\begin{array}{cc}
(M^{\dagger} + \bar{X} N^{\dagger}) (M + X N) & f N^{\dagger} \\
f N & (M + X N) (M^{\dagger} \bar{X} + N^{\dagger})
\end{array}
\right), 
\label{eq:mb}
\\
m_F^2 &=& 
\left(
\begin{array}{cc}
(M^{\dagger} + \bar{X} N^{\dagger}) (M + X N) & 0 \\
0 & (M + X N) (M^{\dagger} \bar{X} + N^{\dagger})
\end{array}
\right).
\label{eq:mf}
\end{eqnarray}
If $f = 0$, the point (\ref{local_vacuum}) becomes a
supersymmetric global vacuum. In this case, all the $\phi$-directions
are not tachyonic and the eigenvalues of $m_B^2 = m_F^2$ are
positive. On the other hand, once we turn on $f \not=0$ terms in the
off-diagonal parts in $m_B^2$, these terms can be considered as
perturbations to the positive definite diagonal parts. 
If the off-diagonal parts are smaller than the diagonal parts, 
the eigenvalues of $m_B^2$ 
can stay in their positive values. 
Especially if we focus on the small $X$ region,
the condition that there are no tachyonic directions is given by $|f N|
\ll |M|^2$ which was employed in \cite{Fe}. 
In addition to a local supersymmetry breaking minimum,
in general, there is also a supersymmetric vacuum in a runaway direction. 
The local minimum is thus a meta-stable vacuum. 

Since the fields $\phi^i$ have mass $m^{ij} \sim M^{ij} + X N^{ij}$ at
the origin of $\phi^i$, given
the energy scale lower than this mass, $\phi^i$ are integrated out and
the effective potential for $X$ is obtained. 
The one-loop effective potential of the pseudo modulus $X$ is given by
the Coleman-Weinberg potential \cite{CoWe}
\begin{eqnarray}
V^{\mathrm{CW}}_{\mathrm{eff}} (X) 
= \frac{1}{64 \pi^2} \mathrm{Tr}
\left[
m_B^4 \log \frac{m_B^2}{\Lambda^2}
- 
m_F^4 \log \frac{m_F^2}{\Lambda^2}
\right],
\label{eq:Coleman-Weinberg}
\end{eqnarray}
where $m_B, m_F$ are bosonic and fermionic mass
matrices and $\Lambda$ is the dynamical cutoff scale. 
The total effective potential $V_{\mathrm{eff}} (X) $ is just the sum of $V_{\mathrm{tree}}$ and
$V^{\mathrm{CW}}_{\mathrm{eff}}$ where $V_{\mathrm{tree}}$ is the
tree-level potential. The one-loop effective potential generically fixes
the modulus $X$.

It was shown that the model (\ref{OR}) can 
dynamically break the $U(1)_R$ symmetry at the supersymmetry 
breaking vacuum by the one-loop corrections 
if there is a field that has R-charge different from 0 or 2 \cite{Sh}. 
Once the parameters $M,N$ and $f$ are chosen appropriately, 
the pseudo modulus $X$ is stabilized at $X\not=0$ giving a $U(1)_R$
breaking vacuum as $X$ has non-zero R-charge.

The model (\ref{OR}) has been generalized to ones with global symmetry and 
with more pseudo moduli \cite{Fe}. In this paper, we focus on the 
 simplest model that has global symmetry $U(N)$ and one modulus $X$
which is discussed in \cite{Fe}. The interaction matrices are given by 
\begin{eqnarray}
M &=& \left(
\begin{array}{cccccc}
0 & 0 & 0 & M_5 & 0 & 0\\
0 & 0 & 0 & 0 & M_7 & 0\\
0 & 0 & 0 & 0 & 0 & M_3\\
M_5 & 0  & 0 & 0 & 0 & 0\\
0 & M_7 & 0 & 0 & 0 & 0\\
0 & 0 & M_3 & 0 & 0 & 0 \\
\end{array}
\right), \quad
N = \left(
\begin{array}{cccccc}
0 & N_5 & 0 & 0 & 0 & 0\\
N_5 & 0 & 0 & 0 & 0 & 0\\
0 & 0 & 0 & N_3 & 0 & 0\\
0 & 0 & N_3 & 0 & 0 & 0\\
0 & 0 & 0 & 0 & 0 & 0\\
0 & 0 & 0 & 0 & 0 & 0\\
\end{array}
\right).
\end{eqnarray}
The superpotential is given by 
\begin{eqnarray}
W &=& f X + X N_5 \phi^a_{(5)} \phi_{(-5)a} + X N_3 \phi^a_{(3)} \phi_{(-3)a} \nonumber \\
&&+ M_7 \phi^a_{(7)} \phi_{(-5)a} + M_5 \phi^a_{(5)} \phi_{(-3)a} + M_3
 \phi^a_{(3)} \phi_{(-1)a}, 
\label{global}
\end{eqnarray}
where $\phi^a_{(7)}, \phi^a_{(5)}, \phi^a_{(3)}, \ (a=1, \cdots, N)$ are fundamental, 
$\phi_{(-5)a}, \phi_{(-3)a}, \phi_{(-1)a}$ are anti-fundamental
representations of $U(N)$ global symmetry and $X$ is a $U(N)$ singlet.
The field $\phi_{(r)}$ has an R-charge $r$.
We have chosen the real parameters $f, N_3, N_5, M_3, M_5$ and $M_7$
to be positive.

As we have discussed, this model breaks supersymmetry at the vacuum 
$\phi_{(r)} = 0$ for all $r$ leaving the flat direction parametrized by $X$. 
In addition to this local minimum, 
there is a runaway direction to the true supersymmetric vacuum,
which is obtained from the vacuum conditions (\ref{cond}). 
It is given by 
\begin{eqnarray}
\begin{aligned}
X =& - \frac{M_5}{N_3} \frac{\phi_5}{\phi_3} \epsilon^{-2}, \quad
\phi_{(3)} = \epsilon \phi_3, \quad 
\phi_{(5)} = \epsilon^{-1} \phi_5, \quad 
\phi_{(7)} = \frac{N_5 M_5}{N_3 M_7} \frac{\phi_5^2}{\phi_3} 
\epsilon^{-3}, \\
\phi_{(-1)} =& \frac{N_3 M_5}{N_5 M_3} \frac{\phi_{-3}^2}{\phi_{-5}} 
\epsilon^{-3}, \quad 
\phi_{(-3)} = \epsilon^{-1} \phi_{-3}, \quad
\phi_{(-5)} = \epsilon \phi_{-5},
\label{eq:runaway1}
\end{aligned}
\end{eqnarray}
where the constants $(f, N_3, N_5, \phi_5, \phi_{-5})$ satisfy the
following conditions,
\begin{eqnarray}
f + 2 N_5 \phi_5 \phi_{-5} = 0, \qquad 
N_3 \phi_{3} \phi_{-3} = N_5 \phi_{5} \phi_{-5}.
\label{eq:runaway2}
\end{eqnarray}
The runaway direction is characterized by a parameter 
$\epsilon \rightarrow 0$.

For later convenience, we re-derive the result obtained in \cite{Fe}
for the allowed region of $U(1)_R$ breaking vacua in the parameter space 
of $f, M$ and $N$ at zero temperature.
Because the effective potential of the model with the $U(N)$ global 
symmetry is just the $N$ copy of the one for the model with $N=1$, 
it is enough to consider $N=1$ case to see the one-loop effective 
potential for $X$.
We investigate the allowed regions for $(M_3/M_5, M_7/M_5)$
with $N_3=N_5=1$ and $f/M_5^2=0.001$. 
The last value is chosen
to satisfy the condition $|M^{-2} f N| \ll 1$ to avoid tachyonic 
directions for small $X$.
The result is given in Fig. \ref{high-t_contour}-(a).

In the next section, we consider the one-loop effective potential 
at finite temperature and study the behavior of the parameter region
that allows $U(1)_R$ breaking vacua.
We will see how this parameter region changes according
to non-zero temperature. 

\section{Finite temperature effects}
The finite temperature contribution in one-loop effective potential 
is given by \cite{DoJa}
\begin{eqnarray}
& & V^{(1)}_B (\phi_c) = \frac{T^4}{2 \pi^2} J_B [m^2_B (\phi_c)/T^2], \label{fpotB}\\
& & V^{(1)}_F (\phi_c) = - 2 \lambda \frac{T^4}{2 \pi^2} J_F [m^2_F
 (\phi_c)/T^2], \label{fpotF}
\end{eqnarray}
where the first is real scalar and the second is fermionic contributions.
Here $T$ is temperature, $\phi_c$ are VEVs at vacua and the fermionic degrees of freedom 
$\lambda$ are given by $\lambda = 1$ (Weyl) or $\lambda = 2$ (Dirac).
The functions $J_B$ and $J_F$ are the so-called bosonic and fermionic thermal
functions given by
\begin{eqnarray}
J_B [m^2 \beta^2] &=& \int^{\infty}_{0} \! d x \ x^2 \log 
\left(1 - e^{- \sqrt{x^2 + \beta^2 m^2}} \right), \label{JB} \\
J_F [m^2 \beta^2] &=& \int^{\infty}_{0} \! d x \ x^2 \log 
\left(1 + e^{- \sqrt{x^2 + \beta^2 m^2}} \right),
\label{JF}
\end{eqnarray}
where $\beta = T^{-1}$. The one-loop effective potential is therefore
given by 
\begin{eqnarray}
V_{\mathrm{eff}} = V_{\mathrm{tree}} + V^{\mathrm{CW}}_{\mathrm{eff}} 
+ V_B^{(1)} + V_F^{(1)}. \label{full}
\end{eqnarray}
In the following, we consider the effective potential (\ref{full}) of
the model (\ref{global}). As discussed in the previous section, since the model
has the $U(N)$ global symmetry, the effective potential (\ref{full}) is just
$N$ copy of the one with $N=1$ case. We therefore consider (\ref{full}) with $N=1$
case. 
Since the high and low temperature expansions are useful to see the behaviour of the potential
analytically and to check the full numerical result, these expansions are 
collected in appendix A.

For all the results in the following, the effective potential (\ref{full}) is
 directly calculated by using Coleman-Weinberg potential
 (\ref{eq:Coleman-Weinberg}) and the  thermal potentials
 (\ref{fpotB}) and (\ref{fpotF}) without any approximation formulas.

\begin{figure}[t]
\begin{center}
\subfigure[$T=0$]
{
\includegraphics[scale=.35]{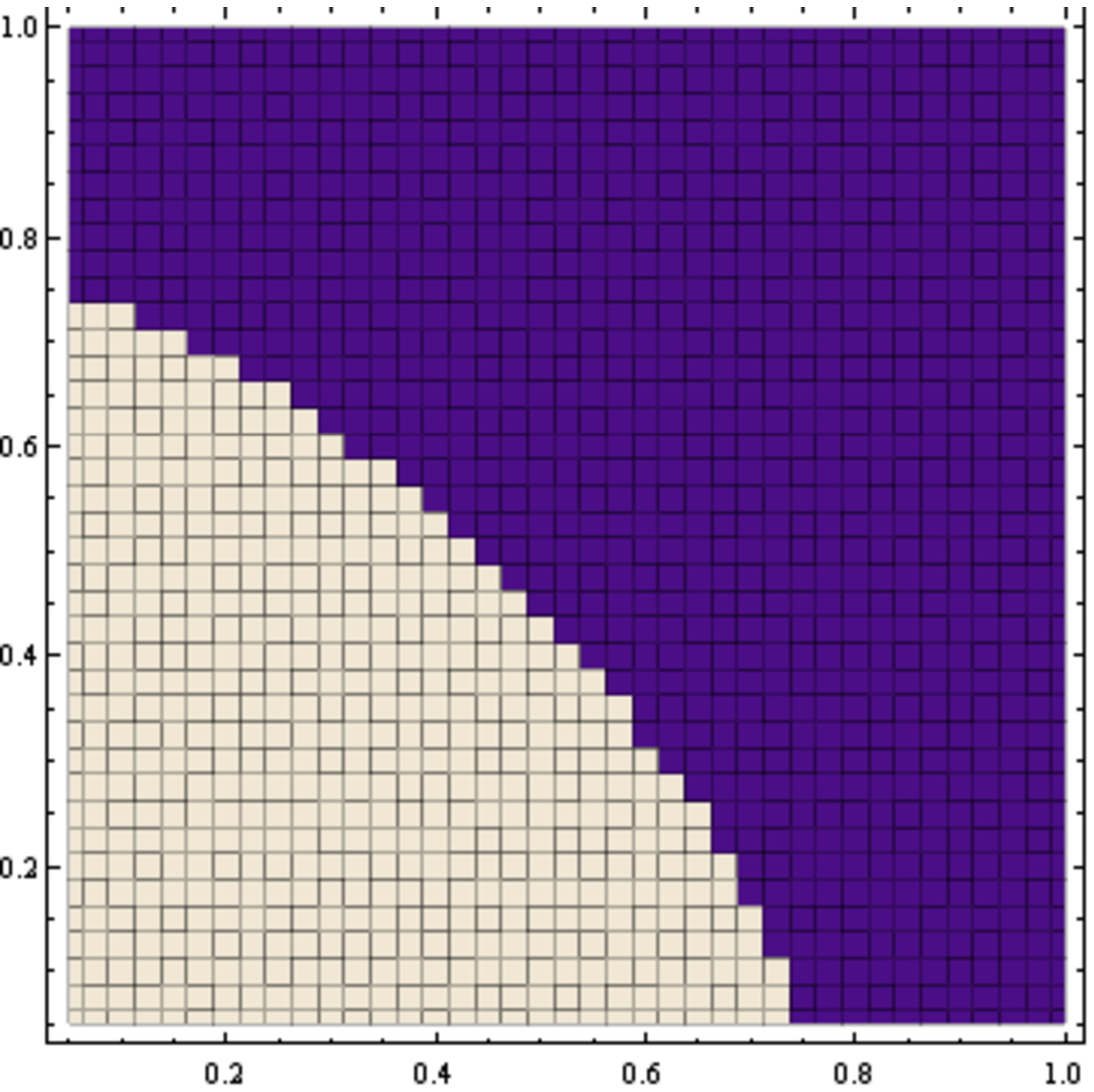}
}
\subfigure[$T=0.2$]
{
\includegraphics[scale=.35]{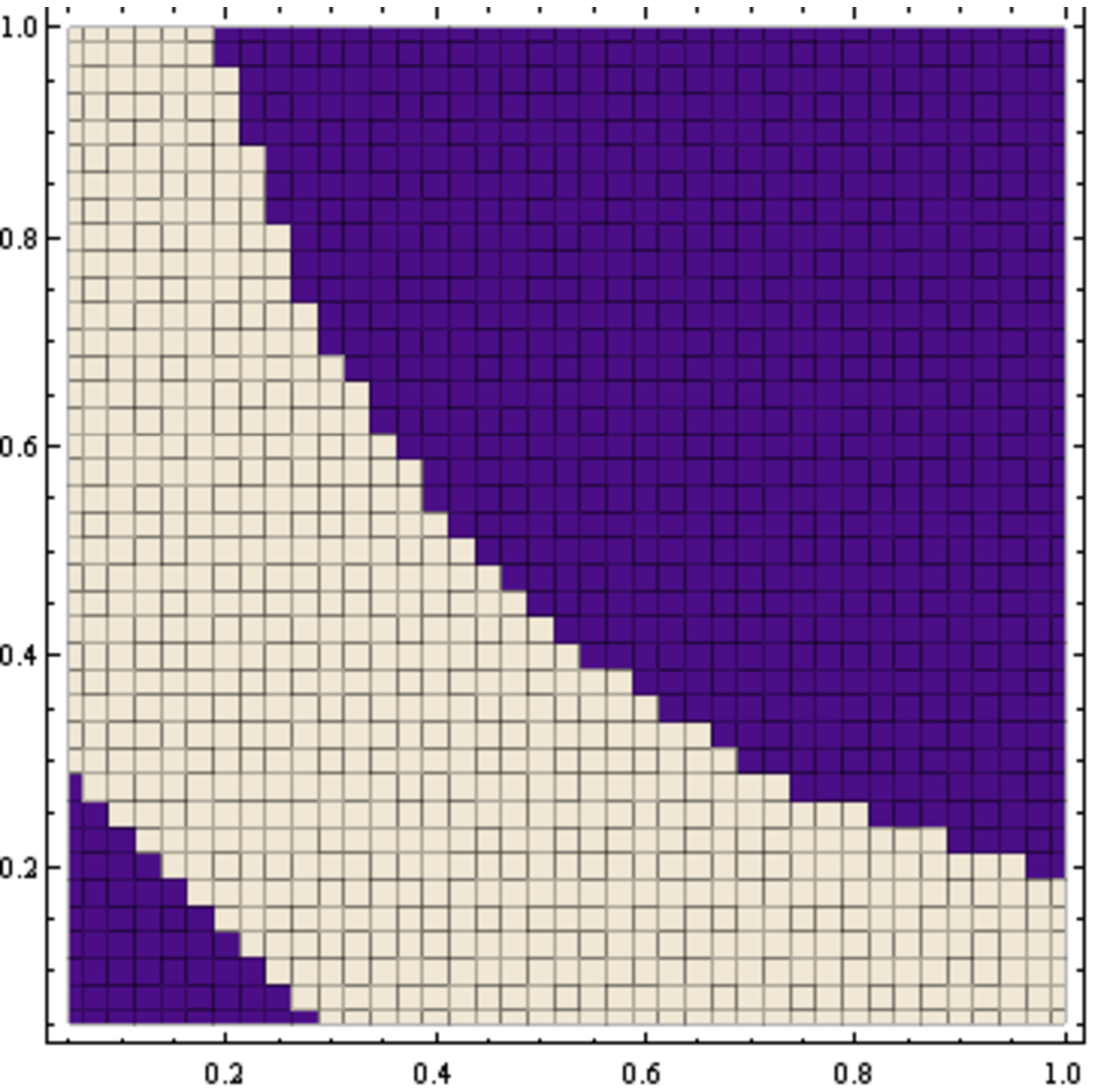}
}
\subfigure[$T=0.3$]
{
\includegraphics[scale=.35]{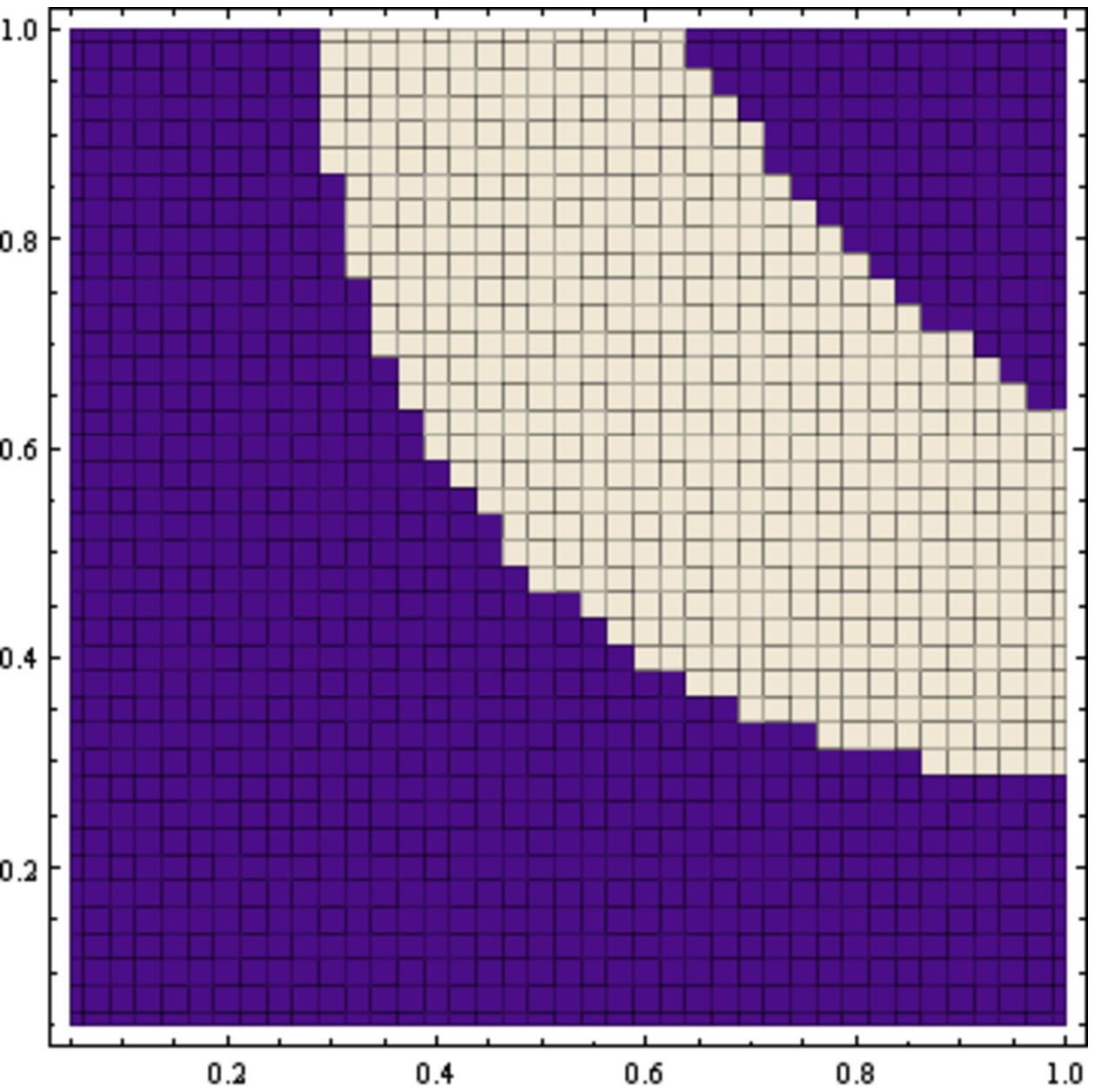}
}
\\
\subfigure[$T=0.5$]
{
\includegraphics[scale=.35]{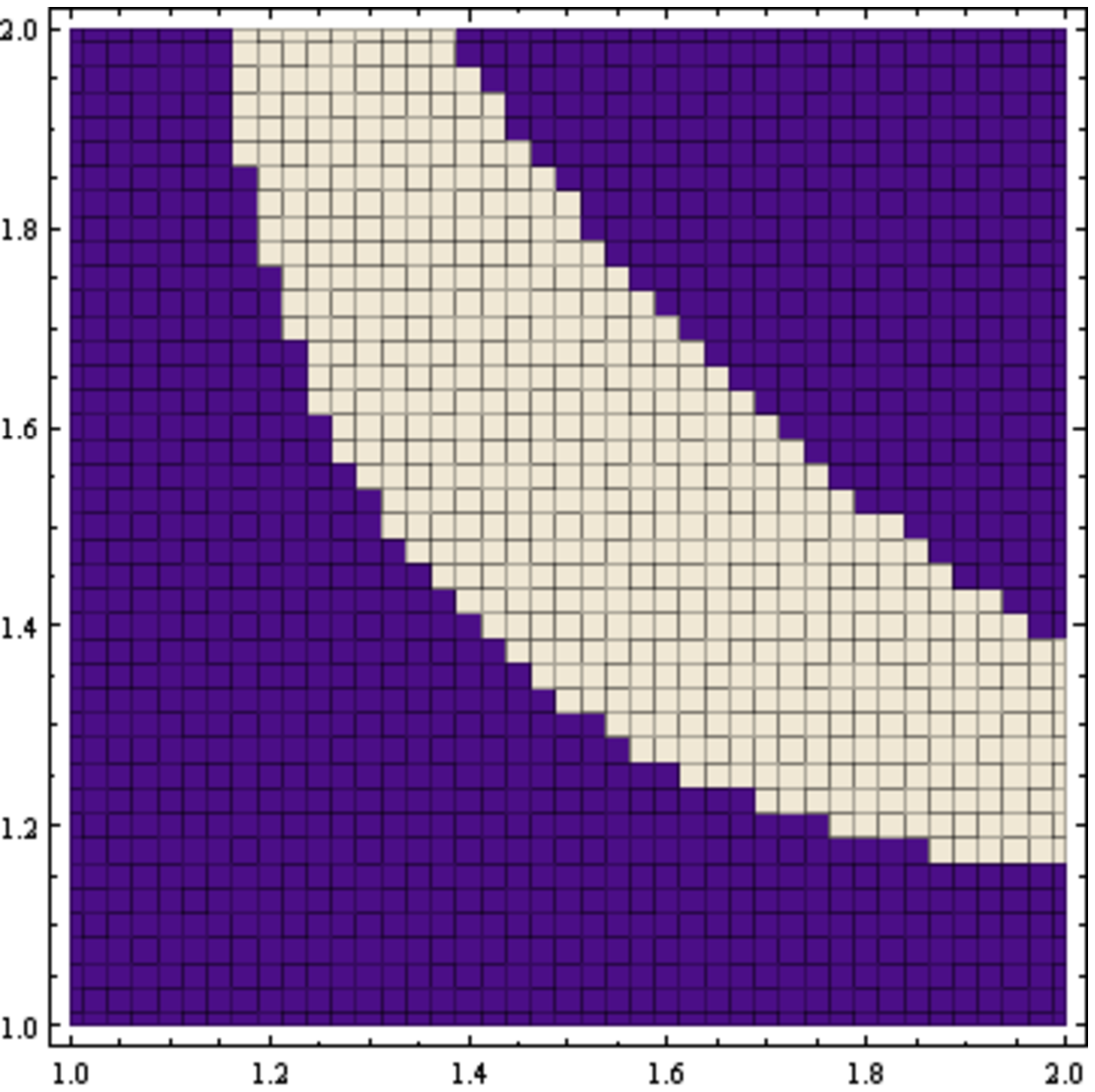}
}
\subfigure[$T=0.7$]
{
\includegraphics[scale=.35]{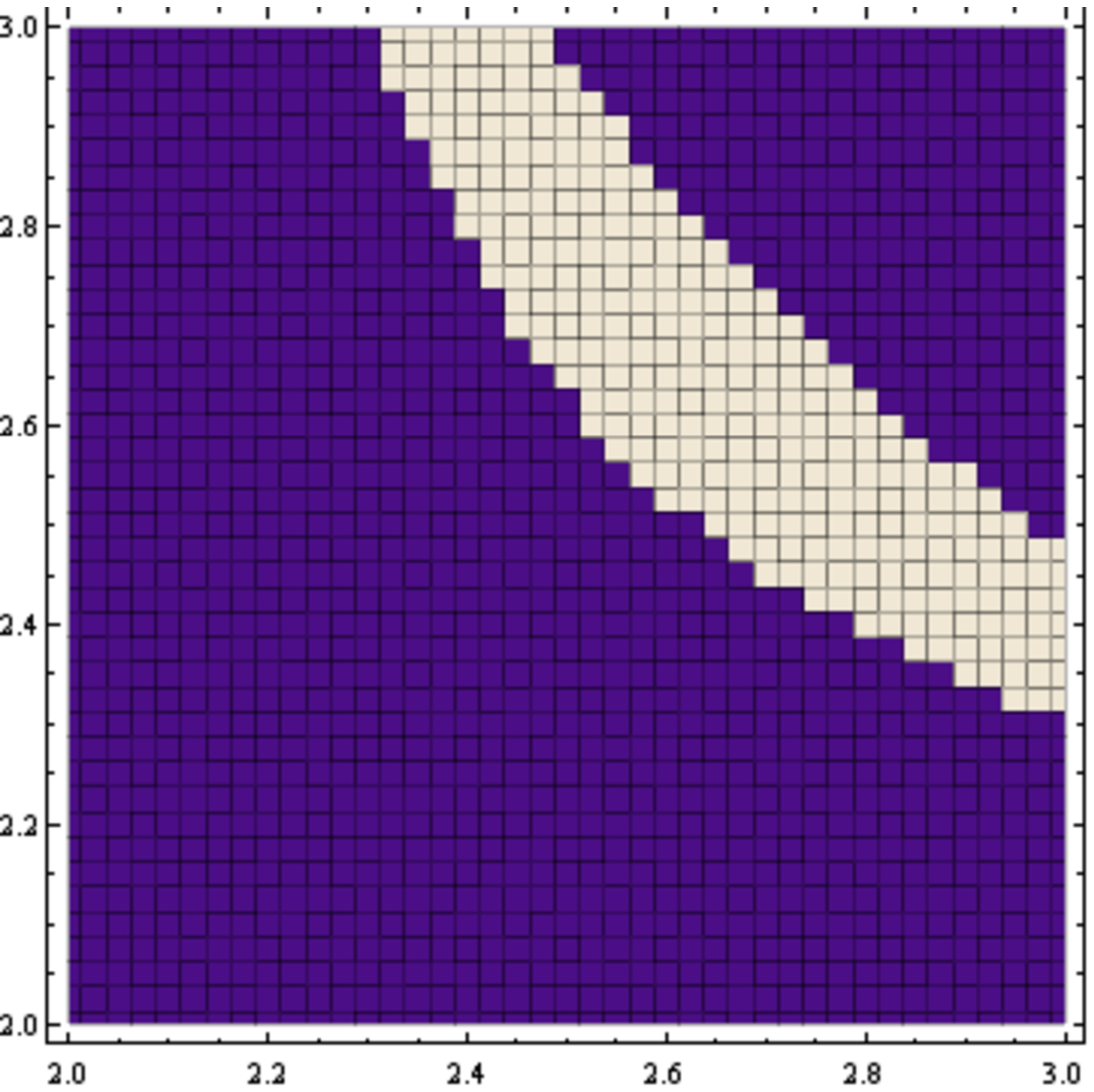}
}
\subfigure[$T=1.0$]
{
\includegraphics[scale=.35]{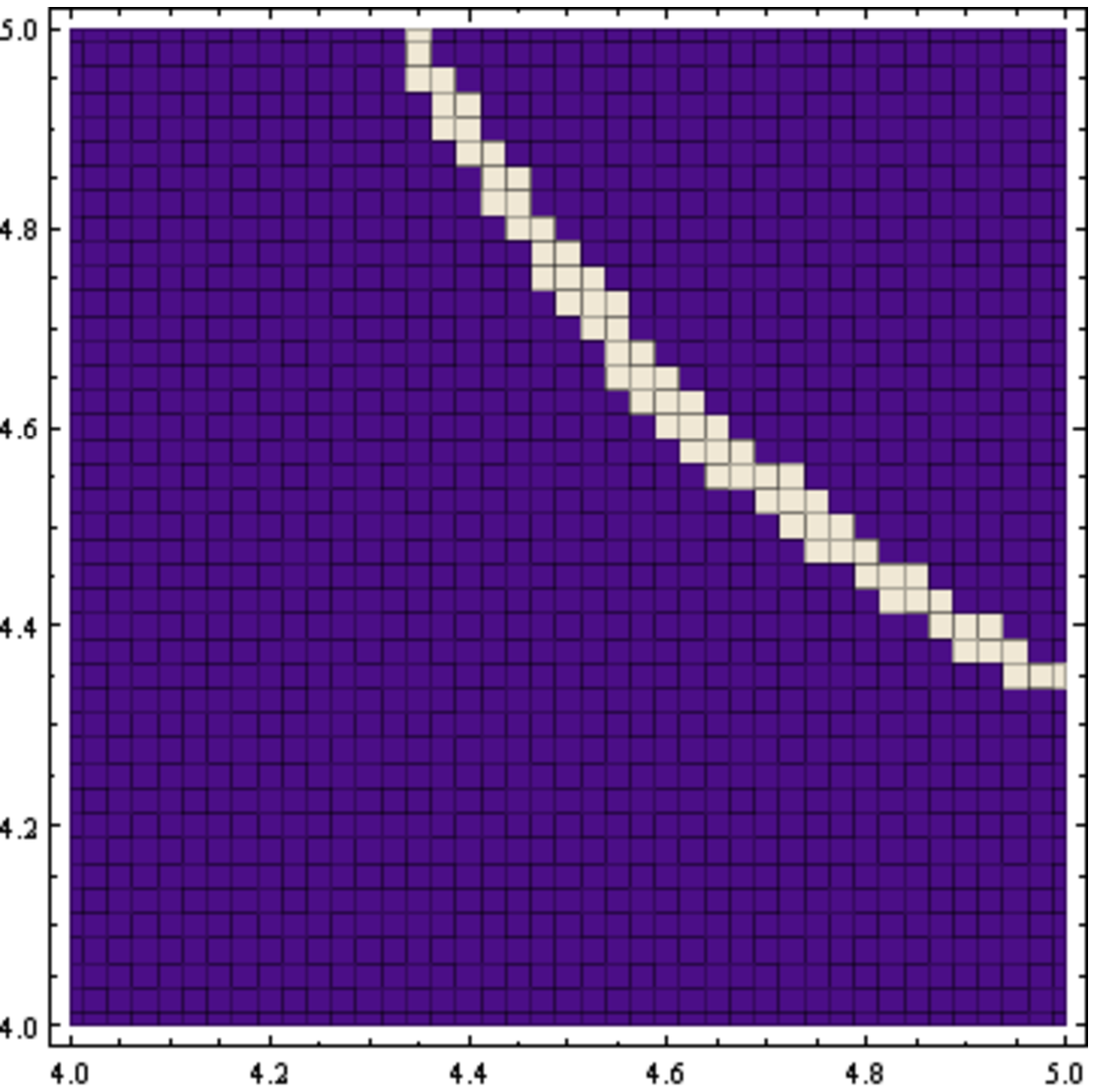}
}
\end{center}
\caption{
Contour plots of the allowed region of the $U(1)_R$ breaking 
against the parameters ($M_3/M_5, M_7/M_5$) for $|X|<3$ with
 temperatures ranging from $T=0$ to $1.0$. 
We choose values of the other parameters as $N_3=N_5=1$, $f/M_5^2=0.001$.
}
\label{high-t_contour}
\end{figure}

We show the parameter region which allows non-supersymmetric
$U(1)_R$ breaking local minimum at temperatures ranging from $T=0$ to 
$T=1.0$ in Fig. \ref{high-t_contour}.
In this analysis, we study the allowed region against the 
parameters $(M_3/M_5, M_7/M_5)$ near the origin of $X$, $|X|\le 3$. 
Their allowed region becomes smaller and the parameters 
$(M_3/M_5, M_7/M_5)$ have larger values, as temperature increases.

The numerical results in Fig. \ref{high-t_contour} enable us to see  
the structure of phase transitions of a given vacuum during the cooling
process. It is seen that at high temperature the $U(1)_R$ symmetry is
generically recovered, which is expected from 
the analytic calculations in Appendix A.
As temperature decreases, with an appropriate choice of parameters, 
a vacuum that preserves the $U(1)_R$ symmetry at high temperature becomes 
a $U(1)_R$ breaking vacuum. 

We note that the numerical result in Fig. \ref{high-t_contour} has
a different behavior from the result of the model in \cite{MoSc} where the
finite temperature effects of the same type of model as (\ref{OR})
without global symmetry have been studied.
In \cite{MoSc}, the contour plots similar to Fig. \ref{high-t_contour} is 
shown and it shows that once certain parameters come into an allowed 
region, they remain in the allowed region until zero temperature. However, in our model, 
some parameters entered into the allowed region
, then it goes out the region as temperature decreases. 
This fact means that there are several phase transitions during the
cooling process.
For instance, for the case with $N_3=N_5=1$ $M_3/M_5=M_7/M_5=0.4$ and 
$f/M_5^2=0.001$, the $U(1)_R$ symmetry vacuum takes place near the origin
around $T\sim 0.25$ and it disappears around $T\sim 0.16$.

Eventually, this parameter choice realizes the $U(1)_R$ symmetry vacuum
at zero temperature, but at lower temperatures, the evolution of vacuum structure is
a bit nontrivial. 
Next, we will show detailed evolutions of the potential at lower 
temperature.

\begin{figure}[t]
\begin{center}
\subfigure[$T=0.2$]
{
\includegraphics[scale=.5]{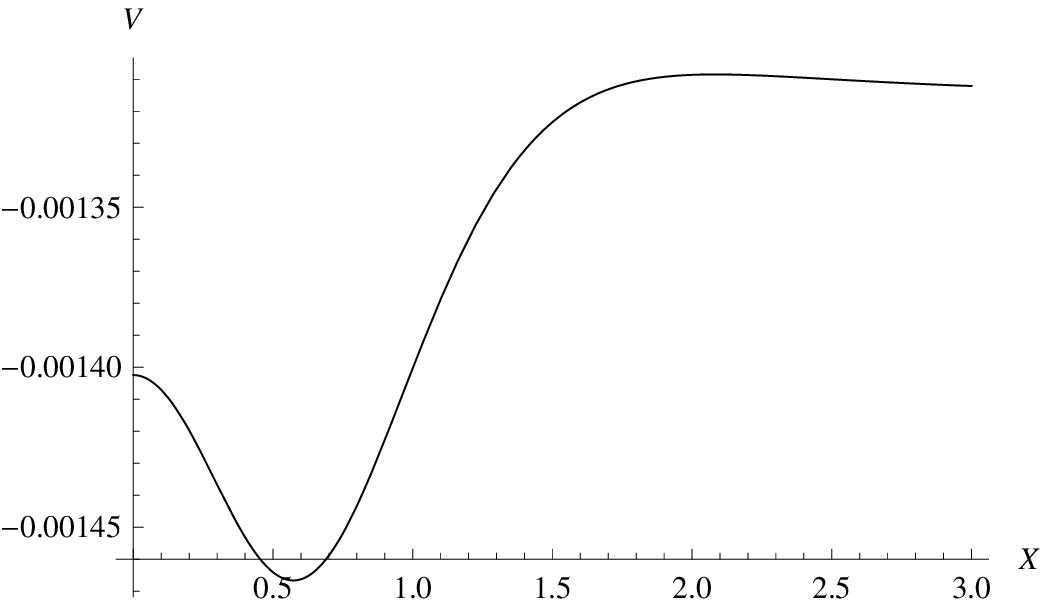}
}
\subfigure[$T=0.18$]
{
\includegraphics[scale=.5]{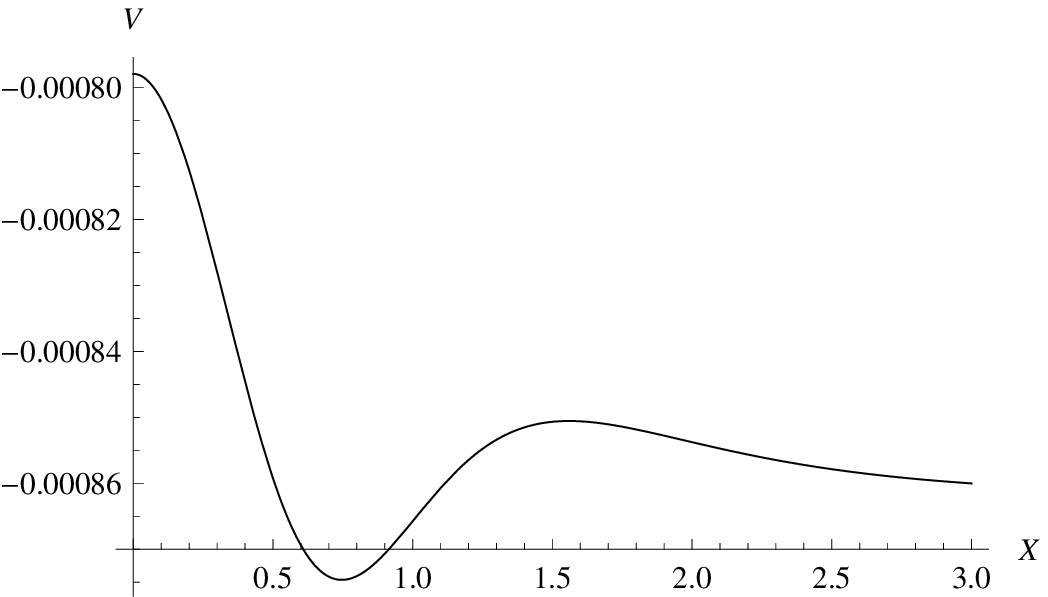}
}
\subfigure[$T=0.16$]
{
\includegraphics[scale=.5]{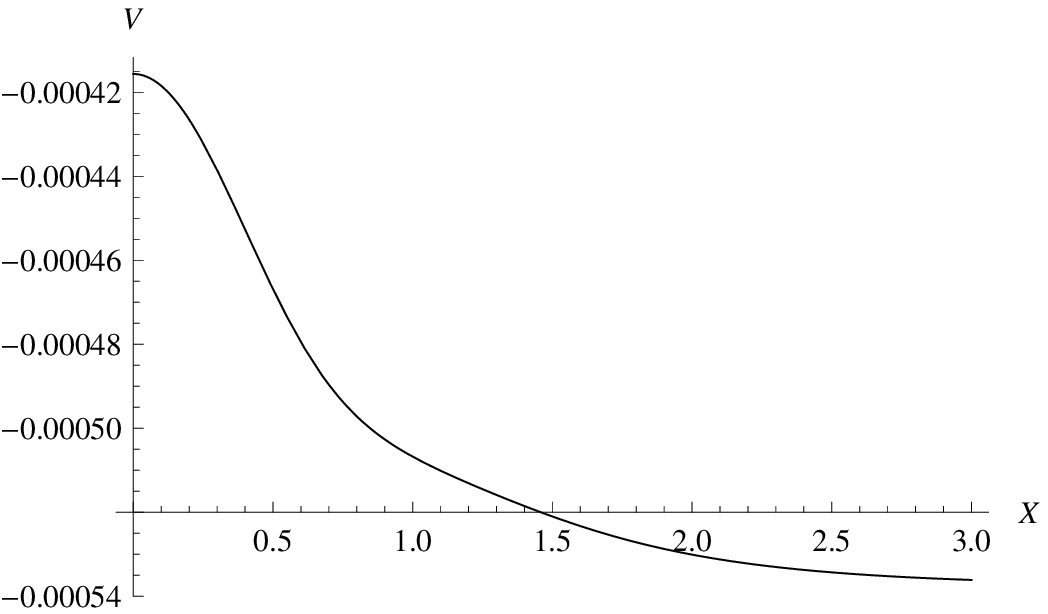}
}
\\
\subfigure[$T=0.02$]
{
\includegraphics[scale=.5]{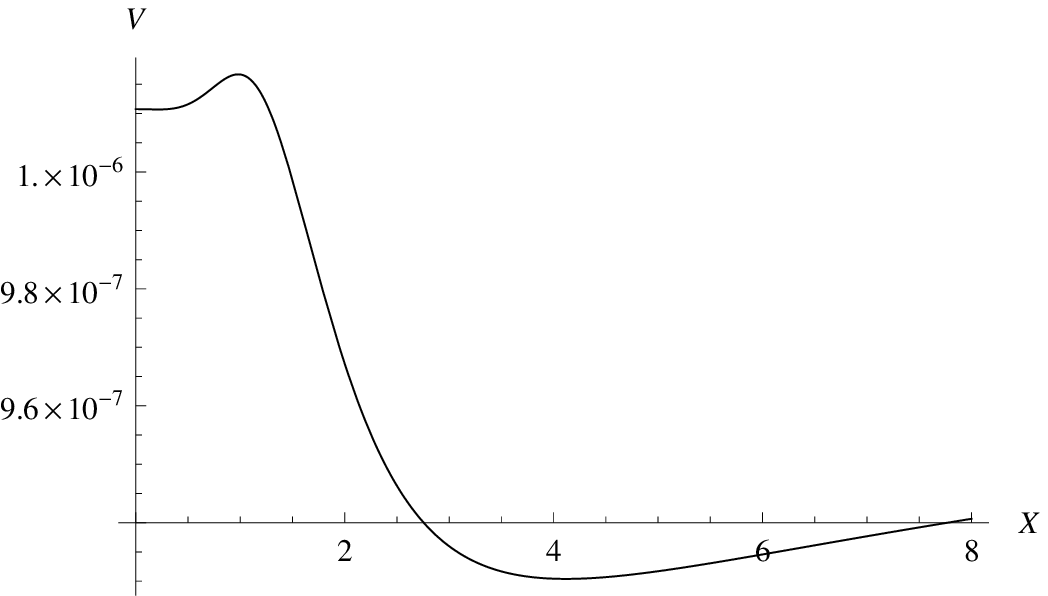}
}
\subfigure[$T=0.015$]
{
\includegraphics[scale=.5]{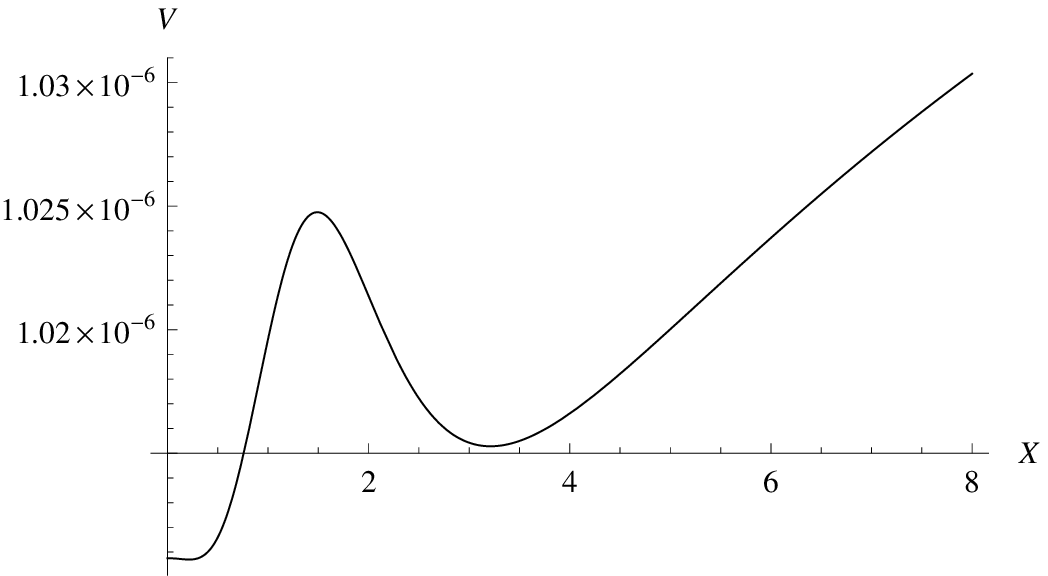}
}
\subfigure[$T=0.01$]
{
\includegraphics[scale=.5]{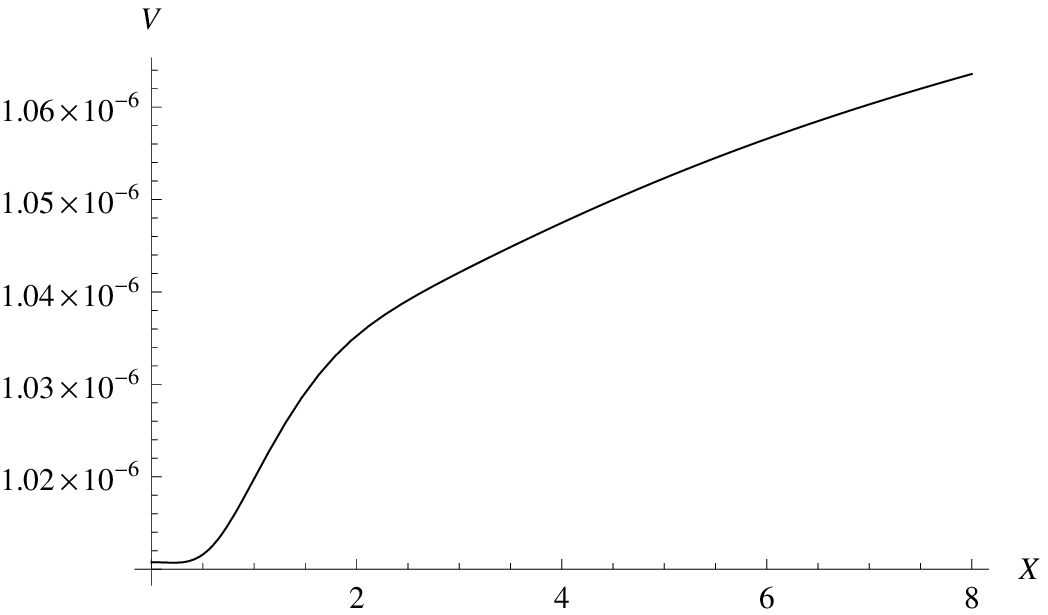}
}
\end{center}
\caption{
Plots of potentials for $X$ with temperatures ranging from $T=0.2$ to
$T=0.01$.
We choose $N_3=N_5=1$, $M_3/M_5=M_7/M_5=0.4$ and $f/M_5^2=0.001$.
In the plots the numerical value is evaluated with $M_5=1$.
}
\label{low-tt}
\end{figure}

Fig. \ref{low-tt} shows the plots of the potentials 
for $X$ with $N_3=N_5=1$, $M_3/M_5=M_7/M_5=0.4$ and $f/M_5^2=0.001$.
As we have discussed, 
there exists a $U(1)_R$ symmetry breaking vacuum near the origin 
for this parameter choice at the temperature around 
$T \sim 0.25$ and the vacuum disappears at a critical temperature as temperature
decreases down to $T\sim 0.16$. 
This behavior can be seen in Fig. \ref{low-tt}-(a), (b) and (c).
Although this $U(1)_R$ breaking vacuum around the origin 
disappeared at the critical
temperature, the $U(1)_R$ breaking minimum re-appears 
at lower temperatures. 
For instance, at $T = 0.16$ we find that there exists a $U(1)_R$ symmetry breaking 
vacuum around $X\sim 13$ which is far from the plot region of Fig. \ref{low-tt}. 
As temperature decreases from $T = 0.16$, this vacuum approaches to the 
origin as can be seen 
around $X\sim 4$ in Fig. \ref{low-tt}-(d). As temperature further
decreases, the $U(1)_R$ symmetry breaking vacuum develops near the
origin again (see Fig \ref{low-tt}-(e)) 
because the Coleman-Weinberg potential becomes dominant and
the extra vacuum far away from the origin disappears as can be seen in Fig. 4-(f).

\section{Runaway direction in finite temperature}
In the previous section we studied the allowed region of the parameter space 
where $U(1)_R$ symmetry is broken. 
One would be afraid about that 
thermal effects on the effective potential may prevent the state
from staying in the vicinity of the meta-stable vacuum
during the cooling process and the system may escape to the runaway
vacuum through the $\phi$-directions.
In this section we study finite temperature effects on 
a potential barrier that intermediates between the meta-stable vacuum
and a runaway direction.

It is not easy to make a thorough investigation how the structure
of the effective potential, which depends not only on $X$ but also
$\phi_{(r)}$, is affected by the change of temperature.
Instead, according to \cite{AbChJaKh1, MoSc}, we choose a particular path
which passes from the meta-stable vacuum to a runaway direction, and
focus on the effective potential of the fixed path.

We choose a path, which is divided into two parts.
The first part of the path includes the meta-stable vacuum while the other includes
a runaway direction. The second path is chosen so that it is free of tachyonic modes.
It is realized by imposing the condition $W^{ijk} W^{\dagger}_{k}=0$ on the bosonic 
mass matrix (\ref{eq:MB}). In other words, it is the condition that the bosonic mass 
matrix coincides with the fermionic one. Then the second part of the path 
sits on the surface determined by the relation $W^{ijk} W^{\dagger}_{k}=0$,
which is two-dimensional subspace in the seven-dimensional space 
spanned by
$(\phi_{(5)},\phi_{(-5)},\phi_{(3)},\phi_{(-3)},\phi_{(7)},\phi_{(-1)},X)$.
The first path is a straight line connecting the meta-stable vacuum and a point on the 
second path. 
The whole path is parametrized by $z$.
The meta-stable vacuum is at $z=0$, and the runaway direction corresponds to $z \to \infty$.

The first part is given by 
\begin{eqnarray}
&&  \phi_{(5)} = - \frac{ \phi_{3} \hat{X_{0}} N_{3}}{M_{5}} z,
 \quad \phi_{(-5)} = \frac{f M_{5}}{2 N_{3} N_{5} \phi_{3} \hat{X_{0}}} z,
 \quad \phi_{(3)} = \phi_{3} z,
 \quad \phi_{(-3)} = - \frac{f}{2 N_{3} \phi_{3}} \zeta, \nonumber \\
&& 
 \phi_{(7)} = \frac{N_{3} N_{5} \phi_{3} \hat{X_{0}}^{2}}{M_{5} M_{7}} z,
 \quad \phi_{(-1)} = \frac{f \hat{X_{0}}}{2 M_{3} \phi_{3}} z,
 \quad X = X_{0} (1-\zeta) + \hat{X_{0}} z \quad (0 \leq z \leq 1).
\label{eq:first part of path}
\end{eqnarray}

The second part of the path is given by
\begin{eqnarray}
&&  \phi_{(5)} = - \frac{ \phi_{3} \hat{X_{0}} N_{3}}{M_{5}} z,
 \quad \phi_{(-5)} = \frac{f M_{5}}{2 N_{3} N_{5} \phi_{3} \hat{X_{0}}}
 \frac{1}{z},
 \quad \phi_{(3)} = \frac{ \phi_{3}}{z},
 \quad \phi_{(-3)} = - \frac{f}{2 N_{3} \phi_{3}} z, \nonumber \\
&&\qquad 
 \phi_{(7)} = \frac{N_{3} N_{5} \phi_{3} \hat{X_{0}}^{2}}{M_{5} M_{7}} z^3,
 \quad \phi_{(-1)} = \frac{f  \hat{X_{0}}}{2 M_{3} \phi_{3}} z^3,
 \quad X = \hat{X_{0}} z^{2}  \quad (1 < z),
\label{eq:second part of path}
\end{eqnarray}
where $X_{0}$ is the VEV of the meta-stable vacuum
at the respective temperatures, and $\hat{X_{0}} = X_{0}(T=0)$.
$\phi_{3}$ is a constant
\footnote{
We choose a simply connected path rather than a smoothly connected
path like in \cite{MoSc}, because even small perturbation around
 (\ref{eq:surfaceeqs}) makes tachyonic modes in the practical calculation.
}.

\begin{figure}[t]
\begin{center}
\subfigure[$T=0, X_0=0.221$]
{
\includegraphics[scale=.55]{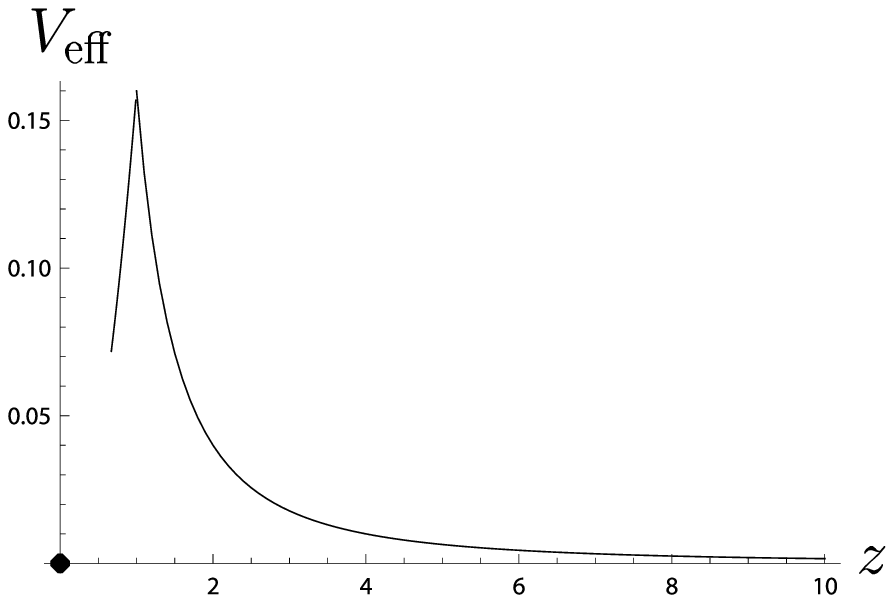}
}
\subfigure[$T=0.25, X_0=0.301$]
{
\includegraphics[scale=.55]{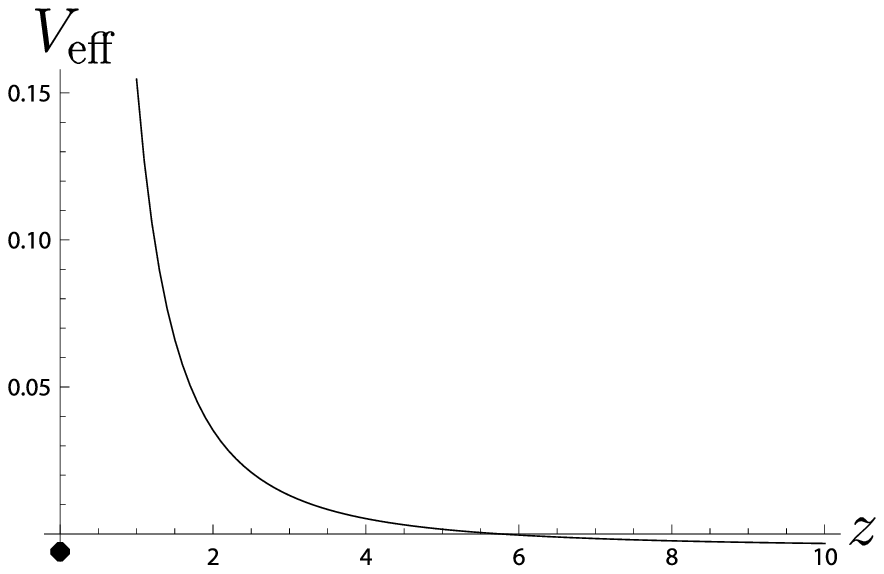}
}
\subfigure[$T=0.5, X_0=0$]
{
\includegraphics[scale=.55]{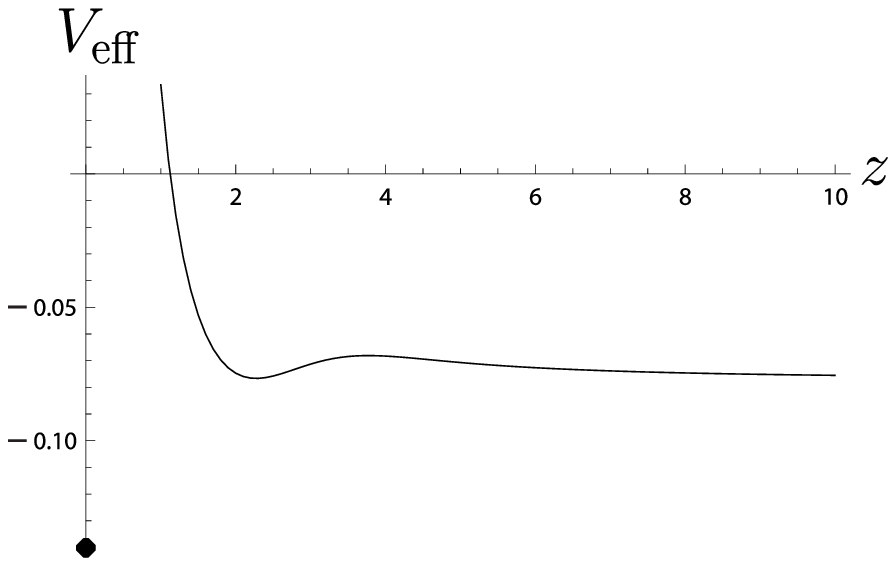}
}
\\
\subfigure[$T=1.0, X_0=0$]
{
\includegraphics[scale=.55]{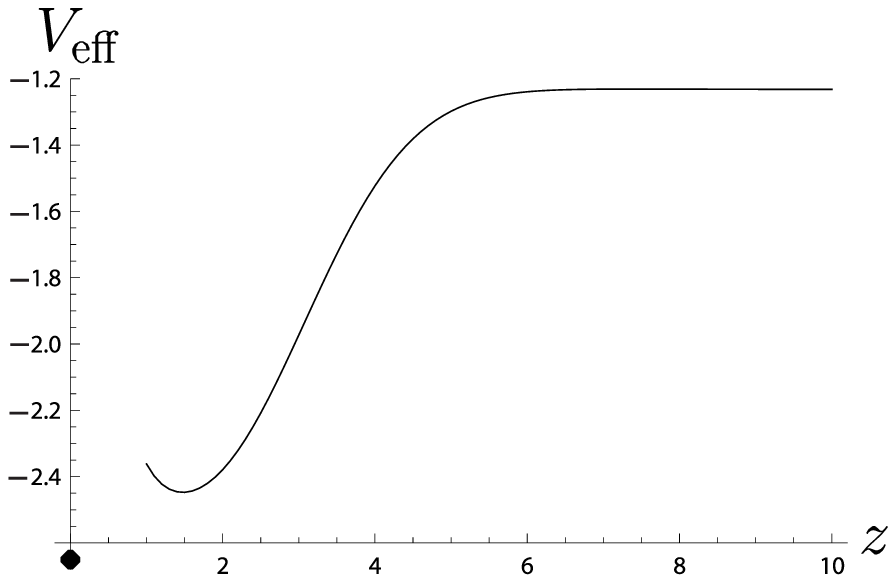}
}
\subfigure[$T=1.5, X_0=0$]
{
\includegraphics[scale=.55]{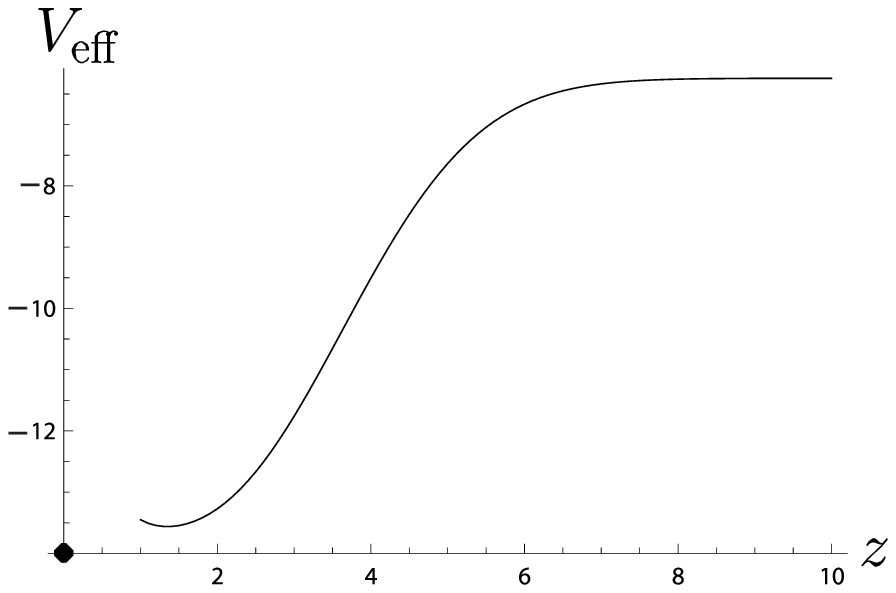}
}
\subfigure[$T=2.0, X_0=0$]
{
\includegraphics[scale=.55]{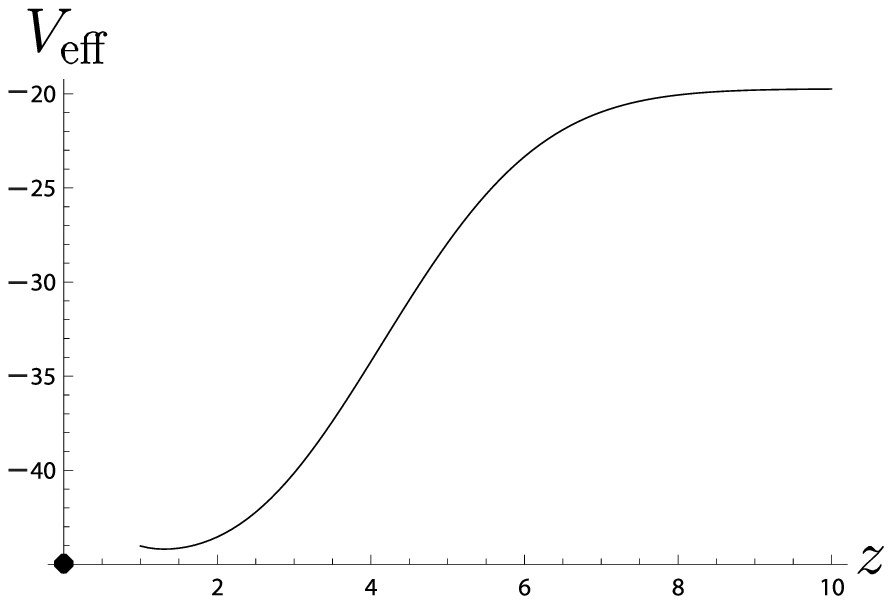}
}
\end{center}
\caption{The thermal effective potential on the path.
The parameters are $M_3/M_5=M_7/M_5=0.4, N_3=N_5=1, f=0.001,$ and
$\phi_3=1/M_5$. The black circles indicate the value of the origin 
($z = 0$) at $X=X_0, \phi =0.$ 
In the interval $0<z<1$ the line is absent if tachyonic modes appear.
In the plots the numerical value is evaluated with $M_5=1$.
}
\label{fig:Low temp runaway}
\end{figure}

The effective potential on the path 
at various temperatures are shown in Fig. \ref{fig:Low temp runaway}. 
The vanishing line
indicates that tachyonic modes appear on the line (we call
this tachyonic line) and states can not stay at that line. 
Therefore the state at the meta-stable vacuum keeps staying around the
origin ($z=0$) and never roll down to the runaway vacuum\footnote{
The absolute value of the effective potential at the origin
is different from that in Fig. \ref{low-tt},
since the size of mass matrices increased by two.
Due to finite temperature effect, 
even a massless field contributes to the effective potential.
This contribution is the constant given by
$- \frac{T^{4}}{2 \pi^{2}} (\frac{\pi^{4}}{45}+\frac{7 \pi^{4}}{360}) \times 2
\approx - 0.41 T^{4}. $
}.

Since $X$ is no longer (pseudo) modulus at  
states with nonzero VEVs for $\phi_{(r)}$,
the mass matrices 
(\ref{eq:mb}) and (\ref{eq:mf}) are not available
to derive the effective potential.
The comprehensive mass matrices 
$M_{B}$ and $M_{F}$ are given in appendix \ref{apx:mass matrix}.
The above condition $W^{ijk} W^{\dagger}_{k}=0$ leads to
\begin{eqnarray}
\begin{aligned}
& \phi_{(5)} = - \frac{N_{3} \phi_{(3)} X}{M_{5}},\,\,
 \phi_{(-5)} = \frac{f M_{5}}{2 N_{3} N_{5} \phi_{(3)} X},\,\,
 \phi_{(-3)} = - \frac{f}{2 N_{3} \phi_{(3)}},\,\,  \\
& \phi_{(7)} = \frac{N_{3} N_{5} \phi_{(3)} X^{2}}{M_{5} M_{7}},\,\,
 \phi_{(-1)} = \frac{f X}{2 M_{3} \phi_{(3)}}.
\label{eq:surfaceeqs}
\end{aligned}
\end{eqnarray}
All VEVs in the mass matrices can be taken to be real by field redefinitions
 without loss of generality.
One can show that (\ref{eq:surfaceeqs}) are equivalent to
(\ref{eq:runaway1}) and (\ref{eq:runaway2})
by replacing $\phi_{(3)} \to \epsilon \phi_{3} $
and $\phi_{(5)} \to \epsilon^{-1} \phi_{5}$.
In fact the tree level potential on the surface (\ref{eq:surfaceeqs}) is written as
\begin{equation}
V_{\rm{tree}} = \frac{f^{2} M_{5}^{2} M_{7}^{2}}{4 N_{3}^{2} N_{5}^{2} |\phi_{(3)}|^{2}|X|^{2} }
+ M_{3}^{2} |\phi_{(3)}|^{2}.
\end{equation}
Taking a limit of $\phi_{(3)} \to 0$, $X \to \infty$ and
$X \phi_{(3)} \to \infty$ at the same time,
$V_{\rm{tree}}$ goes to a supersymmetric global vacuum.

The qualitative behavior of the effective potential in Fig. \ref{fig:Low temp runaway}
is confirmed by an analytic calculation without any specification of parameters.
At high temperature, the expansion (\ref{high-temp}) can be applied.
The effective potential on the surface (\ref{eq:second part of path}) becomes
\begin{eqnarray}
V_{\rm{eff}} &=& V_{\rm{tree}} \nonumber \\
& & - \frac{7}{24}\pi^{2} T^{4} +
 \frac{T^{2}}{16} \left\{ 4 \hat{X_0}^{2}(N_{3}^{2} + N_{5}^{2}) z^{4} 
+ \left( \frac{f^{2}}{ \phi_{3}^{2}}
 + \frac{4 \hat{X_{0}}^{2} \phi_{3}^{2} N_{3}^{2} N_{5}^{2}}{ M_{5}^{2}}
 \right) z^{2}
 + \mathcal{O}(z^{0})
 \right\} \nonumber \\
& & + \frac{T}{2 \pi} \left(
 f \hat{X_{0}} (N_{3}^{2} + N_{5}^{2} ) z^{2} + \mathcal{O}(z^{-2})
 \right) + \cdots,
\end{eqnarray}
where the Coleman-Weinberg potential is absent as mentioned before.
One can see that $V_{\rm eff}$ increases as $\sim z^{4}$ for $z \geq 1$,
and
thus the potential is always minimized at $z=1$ on the second part of the path.
This suggests that, at high temperature,
the stable state exists at least in the region $z \leq 1$.

We summarize the evolution of the vacuum structure on the path 
for the choice of parameters, $N_3=N_5=1, M_3/M_5=M_7/M_5=0.4, f/M_5^2=0.001$
which is considered throughout this paper.
These parameters allow a vacuum with $U(1)_R$ symmetry breaking at zero 
temperature.
At high temperature supersymmetry breaking vacuum sits at the origin 
of the field space $(X,z)$ where $U(1)_R$ symmetry is not broken. 
As temperature decreases, potential barrier between the origin and the
$z$-direction develops and a local vacuum appears along
 $z$-direction (Fig. \ref{fig:Low temp runaway}), which
has a higher potential value than one at the origin. So the origin
remains to be favored as a vacuum.

There is a local vacuum along $z$-direction 
at certain temperatures $T=0.5 \sim 2.0$
which is separated from the meta-stable vacuum by the tachyonic line as
we have mentioned.
Meanwhile, the supersymmetric runaway vacuum develops along
$z \to \infty$ direction around $T=0.5$. 
Although the runaway vacuum may be energetically favoured compared to the
meta-stable vacuum, the local minimum at the origin remains a vacuum
state due to the large potential barrier between the origin and the runaway vacuum.

Simultaneously, the local vacuum at the origin 
of $\phi$ directions ($z=0$) starts to roll down to 
a local vacuum near the origin along the $X$-direction where the $U(1)_R$ 
symmetry is broken.
The large potential barrier between $X$-direction and the 
$\phi$ directions prevents the $U(1)_R$ breaking vacuum 
from decaying to the runaway vacuum and 
keeps it sitting on the meta-stable vacuum.

\section{Conclusions and discussions}
We have studied finite temperature effects on O'Raifeartaigh type models 
with $U(N)$ global symmetry where supersymmetry is spontaneously broken. 
At zero temperature the model has a meta-stable vacuum where $U(1)_R$
symmetry is dynamically broken by the one-loop corrections for appropriate
parameter regions. In addition to a meta-stable vacuum there also exists
a runaway vacuum. We have investigated the finite temperature effects 
in the effective potential of the pseudo modulus field $X$.
In order to consider the finite temperature effects along the runaway 
direction, we chose a particular path so that tachyonic modes can be avoided.

The main features of our study are summarized as follows:
\begin{itemize}
\item
Even though we have taken $N=1$ to study the effective potential, 
the finite temperature effects for 
the superpotential \eqref{global} that allows the global symmetry $U(N)$ is
	 non-trivial. The phase structure during the cooling process is
	 highly dependent on the parameter matrices $M^{ij}$ and $N^{ij}$. 
These matrices can not be chosen freely once one imposes the global
	 symmetries. As we have mentioned in the introduction, the global
	 symmetry is a key point when one considers the mediation of
	 supersymmetry breaking.

\item
Our study reveals the effect of the intermediate and low-temperature region
particularly. 
Many of the preceding analyses do not pay much attention to that region,
however, when one considers the cooling of the universe,
it requires rigorous information in the intermediate and low-temperature
region as well as high-temperature. 
As anticipated from the analytic calculations, the $U(1)_R$ symmetry is
 recovered at high temperature which is the similar result found in the 
O'Raifeartaigh model without global symmetries \cite{MoSc}. 
However, we find non-trivial vacuum structures - the landscape - 
at low temperatures.
Certain parameter regions that do not have the meta-stable vacua
at zero temperature can posses $U(1)_R$ breaking meta-stable vacuum
at finite (low) temperature. 
This structure is not seen in the 
O'Raifeartaigh model without global symmetries \cite{MoSc}.
We obtained this result by performing the full numerical
 analysis without any (high and low temperature) approximations.

\item
To obtain the correct value of the effective potential in runaway direction,
we introduce inherently-tachyon-free path where the bosonic mass matrix 
coincides with the fermionic one.
Another possible way of the path that is used in \cite{MoSc}
 is a smooth connection between
the meta-stable and a runaway vacuum at zero temperature 
with weighting function.
That kind of path is not appropriate for our case
as pointed out in section 4, since in the practical calculation
it makes tachyonic modes on the path and causes a meaningless analyses.
The choice of the path is critical.

\end{itemize}

We would like to make some comments about a previous study.
In the literature \cite{Ka}, thermal effective potential of a class of
O'Raifeartaigh model which includes the model of the present paper
has been analyzed.
The author concludes that within certain high-temperature region
a new meta-stable vacuum that is absent at zero temperature is developed
far away from the origin $X=0$.
This kind of vacuum is not considered in section 4.
The main reason for this is the difference of focusing point and analysis.
In \cite{Ka}, the massive scalar fields $\phi$ are integrated out
to analyze the runaway behaviour $X \to \infty$,
before introducing thermal effects. Then the thermal potential
is given in analytic form by using high-temperature expansion formula.
Meanwhile we keep all the fields explicitly, make the complete mass matrices,
calculate eigenvalues of the mass matrices,
and focus on the effective potential along the fixed path,
particularly not far from the origin of $\phi$
and $X$, by using numerical calculation.

We close this work by pointing out future directions. Here we studied
the hidden sector that the $U(1)_R$ symmetry breaking is spontaneously
broken by the one-loop corrections in ${\cal N}=1$ supersymmetric theory. 
However, in some ${\cal N}=1$ supersymmetric model, especially accompanying
strong gauge dynamics, the perturbative treatment is not available
for all regions of the pseudo flat direction. For instance, in the 
Izawa-Yanagida-Intriligator-Thomas model \cite{IzYa,InTh} where an 
O'Raifeartaigh type sector is dynamically generated by strong gauge dynamics
in the low energy superpotential,
in order to remove degeneracy of pseudo moduli, we have to take account of
quantum corrections for the K\"ahler potential. In general, this is a very 
difficult task since the K\"ahler potential is not holomorphic and thus 
quantum corrections can be estimated at best by perturbative means
only in the ultraviolet (weak coupling) region of the moduli space 
parameterizing the pseudo flat direction which is far from the origin. 
Therefore, the potential behavior in the infrared region remains unclear.
However, in an ${\cal N}=2$ supersymmetric gauge theory one can derive the 
exact low energy effective action as was demonstrated by Seiberg and Witten 
\cite{SeWi1, SeWi2}, using the properties of holomorphy and duality. In 
the framework of ${\cal N}=2$ supersymmetric gauge theory, some  
supersymmetry breaking models with $U(1)_R$ symmetry breaking were considered
\cite{ArOk,ArMoOkSa, OoOoPa, Pa, MaOoOoPa}. 
It would be interesting to study effects of the finite 
temperature in those models (see also \cite{KaPa}).

\subsection*{Acknowledgements}
The work of M.~A. is supported in part by the Research Program MSM6840770029 and by the project of
International Cooperation ATLAS-CERN of the Ministry of Education, 
Youth and Sports of the Czech Republic. 
The work of S.~S. is supported by the Japan Society 
for the Promotion of Science (JSPS) Research Fellowship.

\begin{appendix}

\section{Effective potential at high and low temperatures}
\subsection{High temperature expansion}
In this subsection, we consider series expansions 
of the thermal functions at high temperature.
For $m/T \ll 1$, we have the following expansions,
\begin{eqnarray}
J_B [m^2/T^2] &=& - \frac{\pi^4}{45} + \frac{\pi^2}{12} \frac{m^2}{T^2} 
- \frac{\pi}{6}  \frac{m^3}{T^3}
- \frac{1}{32} \frac{m^4}{T^4} \log \frac{m^2}{a_b T^2} 
\nonumber \\
& & \qquad - 2 \pi^{\frac{7}{2}} \sum_{l = 1}^{\infty}
(-1)^l \frac{\zeta (2 l + 1)}{(l + 1)!} \Gamma \left( 
l + \frac{1}{2} \right) \left( \frac{m^2}{4 \pi^2 T^2} \right)^{l+2},
\end{eqnarray}
and 
\begin{eqnarray}
J_F [m^2/T^2] &=& \frac{7 \pi^4}{360} - \frac{\pi^2}{24} 
\frac{m^2}{T^2} - \frac{1}{32} \frac{m^4}{T^4} \log \frac{m^2}{a_f T^2} 
\nonumber \\
& & \qquad - \frac{\pi^{\frac{7}{2}}}{4} \sum_{l = 1}^{\infty}
(-1)^l \frac{\zeta (2 l + 1)}{( l + 1)!} (1 - 2^{-2l -1}) \Gamma 
\left( l + \frac{1}{2} \right) \left( \frac{m^2}{\pi^2 T^2} \right)^{l 
+ 2},
\end{eqnarray}
where 
$\zeta$ is the Riemann zeta function,
$a_b = 16 \pi^2 \exp (3/2 - 2 \gamma_E)$, $a_f = \pi^2 \exp (3/2 - 2
\gamma_E)$ and $\gamma_E \sim 0.577$ is the Euler-Mascheroni constant.
Therefore the finite temperature part of the effective potential 
(\ref{full}) is given by 
\begin{equation}
V^{(1)} (\phi_c, T) = - \frac{\pi^2 T^4}{90} 
\left(N_s + \lambda \frac{7}{4} N_F  \right)
+ \frac{T^2}{24} \left[ \sum_i^{N_s} m^2_B (\phi_c)_i + \sum_i^{N_F} 
 \lambda m^2_F (\phi_c)_i \right] 
 - \frac{T}{12 \pi} \sum_i^{N_s} m^3_B (\phi_c)_i  + \cdots
\label{high-temp}
\end{equation}
where the dots denote terms that are subleading in $m/T$. 
$N_s, N_F$ are the number of real scalars and Weyl fermions and 
$m_{Bi}, m_{Fi}$ are mass eigenvalues of the boson and fermion mass
matrices. 

In general, it is hard to find the mass eigenvalues analytically.
However, one can extract high temperature effects without solving 
eigenvalue equations explicitly.
At very high temperature $T/m \gg 1$, the $\mathcal{O} (T^2)$ term is 
dominant compare to the $\mathcal{O} (T)$ terms in (\ref{high-temp})
\footnote{In (\ref{high-temp}), $\mathcal{O} (T^4)$
term is just the numerical constant and is irrelevant for analyzing the $X$
dependence.}, the coefficient of $T^2$ in $V^{(1)}$ is given by the
trace of mass matrices $m_B^2$ and $m_F^2$. 
In our model we find 
\begin{eqnarray}
\sum_i {m_{B}^2}_i + \sum_i {m_{F}^2}_i 
= 8 (M_3^2 + M_5^2 + M_7^2 + N_3^2 X^2 + N_5^2 X^2), \label{trace}
\end{eqnarray}
where we have chosen all the parameters real for simplicity.
From the expression of the mass matrices, 
it is easy to see that the result is independent of the parameter $f$. 
The effective potential behaves as
\begin{equation}
V^{(1)} (X,T) \sim T^2 (N_3^2 + N_5^2) X^2 + \mathcal{O} (T). \label{high}
\end{equation}
From this expression, one sees that the modulus $X$ 
tends to be fixed at the origin at high temperature if 
$N_3$ and $N_5$ are not zero. 
Therefore we expect that the region that allows $U(1)_R$ breaking 
vacua is becoming small at high temperature. 
In fact, this behavior can be observed in numerical analysis
of the full effective potential (\ref{full}) 
(see fig. \ref{high-t_contour}).

\subsection{Low temperature expansion}
Let us derive an expression that is useful to study behavior of the effective
potential at low temperature.
For the case $m/T \gg 1$, the bosonic thermal function is expanded as 
\begin{eqnarray}
J_B = \int^{\infty}_{0} \! ds \ s^2 \sum_{n=1}^{\infty} (-1) \frac{X^n
 (s)}{n},
\end{eqnarray}
where $X(s) \equiv e^{- \sqrt{s^2 + m^2 \beta^2}}$.
The integral over $s$ is rewritten by defining new variable $s = m \beta
\sinh t$ as 
\begin{equation}
J_B = \int^{\infty}_0 \! d t \ \sum^{\infty}_{n=1} \frac{1}{n} 
\left(
\frac{m^3}{n^3} \frac{d^3}{d m^3} - \frac{m^3 \beta^2}{n} \frac{d}{d m}
\right) e^{- n m \beta \cosh t}.
\end{equation}
This integration is expressed by the modified Bessel functions:
\begin{equation}
J_B = \frac{m^3 \beta^3}{4} \sum^{\infty}_{n=1} \frac{1}{n} 
\left(
K_1 (nm \beta) - K_3 (n m \beta)
\right),
\label{BesselK_boson}
\end{equation}
where we have used the expression $K_0 (z) = \int_0^{\infty} \! dt \
e^{- z \cosh t}$ and relations
\begin{eqnarray}
{d \over dz} K_0 (z) = - K_1 (z), 
\qquad {d^3 \over d z^3} K_0 (z) = 
- {3 \over 4} K_1 (z) - {1 \over 4} K_3 (z).
\end{eqnarray}
Similarly, we have the fermionic thermal function as
\begin{eqnarray}
J_F = \frac{m^3 \beta^3}{4} \sum^{\infty}_{n=1} \frac{1}{n} (-1)^n
\left(
K_1 (nm \beta) - K_3 (n m \beta)
\right).
\label{BesselK_fermion}
\end{eqnarray}
Let us use the asymptotic expansion of the modified Bessel functions
\begin{eqnarray}
K_{\nu} (z) &\sim & \sqrt{\frac{\pi}{2}} \frac{1}{\sqrt{z}}
e^{-z} \sum^{\infty}_{l=0} \frac{1}{(2z)^l} 
\frac{\Gamma (\nu + l + 1/2)}{l ! \Gamma (\nu - l + 1/2)} \nonumber \\ 
&=& \sqrt{\frac{\pi}{2}} \frac{1}{\sqrt{z}}
e^{-z} \left(
1 + \frac{4 \nu^2 - 1}{8 z} + \cdots
\right), \ (|z| \to \infty),
\end{eqnarray}
where the dots are subleading order in $z^{-l}$. Then, we have
\begin{eqnarray}
J_B (m\beta) \sim - \sqrt{\frac{\pi}{2}} (m \beta)^{\frac{3}{2}}
\mathrm{Li}_{\frac{5}{2}} (e^{- m \beta}).
\end{eqnarray}
Here $\mathrm{Li}_s (z) = \sum_{k=1}^{\infty} \frac{z^k}{k^s}$ is the
polylogarism function and we have used the fact that the dominant
contribution comes from $n=1$ sector in the expansions
(\ref{BesselK_boson}) and (\ref{BesselK_fermion}). 
Again, using the asymptotic behavior $\mathrm{Li}_s (z)
= z + 2^{-s} z^2 + 3^{-s} z^3 + \cdots \ (|z| \to 0)$, we find the
low-temperature expansion of the thermal functions
\begin{eqnarray}
J_{B} (m \beta) &\sim& - \sqrt{\frac{\pi}{2}} ( m \beta)^{\frac{3}{2}} 
\left(
e^{- m \beta} + 2^{-\frac{5}{2}} e^{- 2 m \beta} 
+ 3^{-\frac{5}{2}} e^{- 3 m \beta} + \cdots \right), \\
J_{F} (m \beta) &\sim& - \sqrt{\frac{\pi}{2}} ( m \beta)^{\frac{3}{2}} 
\left(
- e^{- m \beta} + 2^{-\frac{5}{2}} e^{- 2 m \beta} - 3^{-\frac{5}{2}} e^{- 3 
m \beta} + \cdots
\right).
\end{eqnarray}
As we have discussed, at the origin of $\phi^i$, the mass for $\phi^i$
is given by $m \sim M + N X$. Therefore we expect that the finite
temperature part of the effective potential at low temperature behaves like
\begin{eqnarray}
V^{(1)} (X,T) \sim - T^4 {\rm Tr}
\left\{((M+N X) T^{-1})^{\frac{3}{2}} e^{-(M+N X) T^{-1}}\right\}.
\end{eqnarray}
This expression would be useful for a calculable model to see 
the thermal evolution of vacuum analytically at low temperature. 
In our case, since analytic forms of eigenvalues are complicated, 
it is difficult to see the evolution analytically. However, the 
low temperature expansion is still useful for numerical calculations. 
It needs shorter time to evaluate the effective potential compare to 
the full thermal functions \eqref{JB}, \eqref{JF}. 
One can use the low temperature expansion to check the full result 
in the numerical analysis in section 3.

\section{Mass matrices}
\label{apx:mass matrix}
The general mass matrices are given as follows.
\begin{eqnarray}
M^{2}_{\rm{B}} &=& \left( 
\begin{array}{cc}
W^{\dagger}_{ik} W^{kj} & W^{\dagger}_{ijk} W^{k} \\
W^{ijk} W^{\dagger}_{k} & W^{ik} W^{\dagger}_{kj} 
\end{array}
\right),
 \label{eq:MB}
\\
M^{2}_{\rm{F}} &=& \left( 
\begin{array}{cc}
W^{\dagger}_{ik} W^{kj} & 0 \\
0 & W^{ik} W^{\dagger}_{kj} 
\end{array}
\right),
 \label{eq:MF}
\end{eqnarray}
where
\begin{eqnarray}
& & W^{\dagger}_{ik} W^{kj} = (W^{ik} W^{\dagger}_{kj})^{\dagger} =
 \nonumber \\
& & \left(
\begin{array}{c|c|c|c|c|c|c}
\scriptstyle{M_5^2+N_5^2(|X|^2} & \scriptstyle{N_5^2\phi_5 \phi_{-5}^{\dagger}}
 &  \scriptstyle{N_3(M_5 X} & \scriptstyle{N_3 N_5 \phi_3 \phi_{-5}^{\dagger}}
 & \scriptstyle{ M_7 N_5 X^{\dagger}} & 0
 & \scriptstyle{ M_5 N_3 \phi_3 + N_5^2 X^{\dagger} \phi_5} \\
 \scriptstyle{ +|\phi_{-5}|^2)} & 
 &  \scriptstyle{ +N_5 \phi_{-3} \phi_{-5}^{\dagger})} &
 & &
 & \\ \hline
\scriptstyle{N_5^2 \phi_{-5} \phi_5^{\dagger}} &  \scriptstyle{M_7^2 + N_5^2(|X|^2 }
 &  \scriptstyle{N_3 N_5 \phi_{-3} \phi_5^{\dagger}} & \scriptstyle{ N_5(M_5 X^{\dagger}}
 & 0 & 0
 &  \scriptstyle{N_5^2 \phi_{-5} X^{\dagger}} \\
 & \scriptstyle{ + |\phi_5|^2)}
 & & \scriptstyle{  + N_3 \phi_3 \phi_5^{\dagger})}
 & &
 & \\ \hline
\scriptstyle{N_3 (M_5 X^{\dagger} } & \scriptstyle{ N_3 N_5 \phi_5 \phi_{-3}^{\dagger} }
 & \scriptstyle{ M_3^2 + N_3^2(|X|^2 } &  \scriptstyle{ N_3^2 \phi_3 \phi_{-3}^{\dagger} }
 & 0 & 0 
 &  \scriptstyle{N_3^2 X^{\dagger} \phi_3} \\
 \scriptstyle{+ N_5 \phi_{-5} \phi_{-3}^{\dagger})} &
 & \scriptstyle{ + |\phi_{-3}|^2)} &
 & &
 & \\ \hline
\scriptstyle{N_3 N_5 \phi_{-5} \phi_3^{\dagger}} &  \scriptstyle{N_5(M_5 X}
 &  \scriptstyle{ N_3^2 \phi_{-3} \phi_{3}^{\dagger} } &  \scriptstyle{M_5^2 + N_3^2(|X|^2} 
 & 0 &  \scriptstyle{ M_3 N_3 X^{\dagger} }
 & \scriptstyle{ M_5 N_5 \phi_{-5} + N_3^2 X^{\dagger} \phi_{-3}} \\
 & \scriptstyle{ + N_3 \phi_5 \phi_3^{\dagger} )}
 & & \scriptstyle{ + |\phi_3|^2)}
 & &
 & \\ \hline
\scriptstyle{M_7 N_5 X} & 0
 & 0 & 0
 &  \scriptstyle{M_7^2} & 0
 &  \scriptstyle{M_7 N_5 \phi_5} \\
 &
 & &
 & &
 & \\ \hline
0 & 0
 & 0 &  \scriptstyle{M_3 N_3 X}
 & 0 &  \scriptstyle{M_3^2}
 &  \scriptstyle{M_3 N_3 \phi_{-3}} \\
 &
 & &
 & & 
 & \\ \hline
\scriptstyle{M_5 N_3 \phi_3^{\dagger} + N_5^2 X \phi_5^{\dagger}} &  \scriptstyle{N_5^2 X \phi_{-5}^{\dagger} }
 & \scriptstyle{N_3^2 X \phi_3^{\dagger}} & \scriptstyle{M_5 N_5 \phi_{-5}^{\dagger}}
 & \scriptstyle{M_7 N_5 \phi_5^{\dagger}} & \scriptstyle{M_3 N_3 \phi_{-3}^{\dagger}}
 & \scriptstyle{N_5^2(|\phi_{3}|^2+|\phi_{5}|^2} \\
 &
 & & \scriptstyle{ + N_3^2 X \phi_{-3}^{\dagger}}
 & &
 & \scriptstyle{  +|\phi_{-3}|^2+|\phi_{-5}|^2)}
\end{array}
\right),
\nonumber \\
\end{eqnarray}

\begin{eqnarray}
& & W^{\dagger}_{ijk} W^{k} = ( W^{ijk} W^{\dagger}_{k} )^{\dagger} =
 \nonumber \\
& &  \left(
\begin{array}{c|c|c|c|c|c|c}
 0 & \scriptstyle{N_5(f+N_3 \phi_{3} \phi_{-3}}
 & 0 & 0
 & 0 & 0
 & \scriptstyle{ N_5(N_5 X \phi_5} \\
 & \scriptstyle{ +N_5 \phi_{5} \phi_{-5}) }
 & &
 & &
 & \scriptstyle{ + M_7 \phi_7) } \\ \hline
\scriptstyle{N_5(f+N_3 \phi_{3} \phi_{-3}} & 0
 & 0 & 0
 & 0 & 0
 & \scriptstyle{ N_5(N_5 X \phi_{-5}} \\
 \scriptstyle{+N_5 \phi_{5} \phi_{-5})} & 
 & &
 & &
 & \scriptstyle{ + M_5 \phi_{-3}) }\\ \hline
0 & 0
 & 0 & \scriptstyle{N_3(f+N_3 \phi_{3} \phi_{-3}}
 & 0 & 0
 & \scriptstyle{ N_3(N_3 X \phi_3} \\
 &
 & & \scriptstyle{ + N_5 \phi_{5} \phi_{-5})}
 & &
 & \scriptstyle{ + M_5 \phi_5) }\\ \hline
0 & 0
 & \scriptstyle{N_3(f+N_3 \phi_{3} \phi_{-3}} & 0
 & 0 & 0
 & \scriptstyle{ N_3(N_3 X \phi_{-3}} \\
 &
 & \scriptstyle{ + N_5 \phi_{5} \phi_{-5}) } &
 & &
 & \scriptstyle{ + M_3 \phi_{-1} ) } \\ \hline
0 & 0
 & 0 & 0
 & 0 & 0
 & 0 \\
 &
 & &
 & & 
 & \\ \hline
0 & 0
 & 0 & 0
 & 0 & 0
 & 0 \\
 &
 & &
 & & 
 & \\ \hline
\scriptstyle{ N_5(N_5 X \phi_5} & \scriptstyle{ N_5(N_5 X \phi_{-5}}
 &  \scriptstyle{ N_3(N_3 X \phi_3 } & \scriptstyle{ N_3(N_3 X \phi_{-3}} 
 & 0 & 0
 & 0 \\
\scriptstyle{ + M_7 \phi_7) } & \scriptstyle{ + M_5 \phi_{-3}) }
 & \scriptstyle{ + M_5 \phi_5) } & \scriptstyle{ + M_3 \phi_{-1}) }
 & &
 &
\end{array}
\right).
\nonumber \\
\end{eqnarray}

$X$ and $\phi_r$ are VEVs of $X$ and $\phi_{(r)}$ respectively.

\end{appendix}


\end{document}